\documentclass[12pt]{article}
\usepackage{times}
\usepackage{bm}
\usepackage{natbib}
\usepackage{graphicx}
\usepackage{amsmath, amssymb}

\newcommand{\bP}{\mathbb{P}}
\newcommand{\cS}{{\cal S}}
\newcommand{\cs}{{s}}
\newcommand{\cC}{{\cal C}}
\newcommand{\cG}{{\cal G}}
\newcommand{\tP}{\text{pr}}

\begin{document}

\begin{center}
  { \Large Computing the likelihood of sequence segmentation under
    Markov modelling}

  Laurent GU\'EGUEN 

  Universit\'e de Lyon; universit\'e Lyon 1; CNRS; UMR 5558,
  Laboratoire de Biom\'etrie et Biologie \'Evolutive, 43 boulevard du
  11 Novembre 1918, Villeurbanne F-69622, France.

  gueguen@biomserv.univ-lyon1.fr
\end{center}

\newpage

\begin{abstract}
  I tackle the problem of partitioning a sequence into homogeneous
  segments, where homogeneity is defined by a set of Markov models.
  The problem is to study the likelihood that a sequence is divided
  into a given number of segments. Here, the moments of this
  likelihood are computed through an efficient algorithm. Unlike
  methods involving Hidden Markov Models, this algorithm does not
  require probability transitions between the models. Among many
  possible usages of the likelihood, I present a maximum \textit{a
  posteriori} probability criterion to predict the number of
  homogeneous segments into which a sequence can be divided, and an
  application of this method to find CpG islands.
\end{abstract}

\newpage

\section{Introduction}

An important element in analysing a sequence of letters is to find out
whether the sequence has a structure, and if so, how it is structured.
Usually, looking for structure in a sequence implies a partition -- or
segmentation -- in which each segment can be considered
``homogeneous'', on the basis of a specific criterion. There are two
main approaches to tackle this problem~\citep{BM98}.

\medskip

A commonly used methodology is to model the sequence with Markov
models. A Markov model gives, for each word of a given length, the
probabilities of letters conditionally following this word -- called
emission probabilities. The likelihood of a segment of letters is the
product of these probabilities at all the positions of the segment.
Various models give different likelihoods for a given segment, some of
them greater than others. Looking for a segmentation of a sequence
means dividing it into segments, so that a model chosen as the best
from amongst a set of models is attributed to each segment. One way to
study the structure of a sequence is to analyse the set of its
segmentations.

To make this task possible, the set of models is usually organized to
form a Markov meta-model in which there are transition probabilities
between the models. This is known as a Hidden Markov Model (HMM). In
this context, the models are usually called states, but for the sake
of consistency I keep the same vocabulary as before. In an HMM, a run
of models is a Markov process with a probability, and, given a run of
models, the sequence has a likelihood. If a segment is defined as a
range of positions modelled by a unique model, it is possible to
compute the probability of a segmentation given the sequence and the
HMM. As this method permits efficient (i.e.~with linear complexity)
algorithms for sequence analysis and partitioning~\citep{Rab89}, it is
used in numerous applications, for example in bioinformatics
\citep{Chu89, BCHM94, LB98, PG99, NBMH02, BH04} and in speech
recognition~\citep{ODK96}.

However, since in an HMM the chain of the models is markovian, the
lengths of the segments defined by the models are expected to follow
geometric laws, which may be a false hypothesis for real data
segments. Various solutions have been proposed to overcome this
problem, such as using semi-Markov chains~\citep{Gue05a} or
macro-states~\citep{EM02}, but in fact they make the modelling task
more complex, since more parameters are used to obtain a better
modelling of the lengths of the segments. Moreover, in the problem of
sequence segmentation using a set of models, the inter-model
transition probabilities used in an HMM correspond to an \textit{a
priori} on the distributions of the segments, and are superfluous
parameters if we consider that the models themselves should be
sufficient to segment the sequence, as in the approach described
below. Finally, in an HMM, the models modelling and the length
modelling can be seen as two competing modellings, because in the
parts of the sequence where the models do not discriminate clearly,
the length parameters will have a predominant influence. This is even
more problematic when the lengths of the real segments are very
different along the sequence.

\medskip 

A way to avoid these ``extra'' parameters is to establish a
homogeneity criterion for a segment (such as the variance of its
composition, or its maximum likelihood given specific models), and to
determine a set of segments that divide the sequence and minimize --
or maximize -- this criterion. This problem -- also known as the
changepoint problem -- can be solved by an optimal
algorithm~\citep{Bel61}, but its time-complexity is quadratic with the
length of the sequence, which prohibits the analysis of very long
sequences. Alternatively, this problem can be tackled linearly using
hierarchical segmentation \citep{LBHG02, Li01a}, or with
approximations about the limits of the segments~\citep{BH93, BBM00},
but these approaches do not ensure that the best partition is found.
Moreover, when the homogeneity criterion is monotonous with the number
of segments (such as the maximum likelihood of markovian processes),
these methods need an additional criterion to stop the segmentation
process. For each number of segments, the calculation of the criterion
is based on the built partition and is very dependent on the choice of
this partition. Without a stopping criterion, these methods produce
multi-level descriptions of the structure of the sequences that may be
quite interesting, but I am not aware of any practical usage of such
sets of segmentations.

\medskip

Between those two approaches, I described in ~\citep{Gue01} an
algorithm -- known as MPP, or Maximal Predictive Partitioning -- that
computes the most likely segmentation of a sequence in $k$ segments
given a set of Markov models. This algorithm is optimal and has a
time-complexity linear with the length of the sequence. As with the
previous segmentation methods, it provides a multi-level description
of the structure of a sequence, and it needs an additional criterion
to select the ``best'' partition, such as the number of segments.

\medskip

Bayesian methods are a different approach to work on sequence
segmentation, since they propose to simulate the \emph{a posteriori}
distribution of the segmentations of a sequence, given a
criterion~\citep{LL99, SKM02, MRGRT01, Kei06}. Even though they do not
construct the best segmentation, they indicate the relative
significance of the segmentations, and the structuring of the
sequence. Nonetheless, as the set of segmentations is very large, the
convergence of the simulated distribution towards the right one can be
extremely slow.

\medskip

I would now like to look at the problem of estimating the structuring
of a sequence given a set of Markov models. In contrast to the
situation for an HMM, I do not want to put any constraint on the
transitions between the models. This article presents an algorithm
that computes the moments of the likelihood of a sequence under the
set of all partitions with a given number of segments. The maximum of
this likelihood was already computable with the MPP
algorithm~\citep{Gue01}. Since the time-complexity of this new
algorithm is linear with the length of the sequence, it can also be
applied to very long sequences.

The distribution of this likelihood may be useful for many statistical
analyses of sequences, for example in an HMM modelling to test for the
relevance of inter-model transition probabilities, or in a change
point problem to test the significance of partitions and stop the
partitioning, or in a bayesian approach to perform more efficient
simulations of the \textit{a posteriori} distribution of the
segmentations of a sequence. As an example, I propose a maximum
\emph{a posteriori} estimator of the numbers of segments in a
sequence.

\newpage

\section{Method\label{sec:method}}

\subsection{Computing the likelihood of the sequence}

The method computes the moments of the likelihood that a sequence is
partitionned in exactly $k$  segments given a set of Markov models.
The algorithm that is presented permits the computation of the mean of
this distribution. Generalizing this to the computation of all moments
is straightforward.

\medskip

First, we introduce some notations and concepts.

The studied sequence, $\cS$, consists of letters, and has a length
$l$. For all $i \in [0,l-1]$, we denote by $\cs_i$ the $i$-th letter
of $\cS$, and $\cS_i$ the segment of $\cS$ from position $0$ to
position~$i$, inclusive. $\cS=\cS_{l-1}$.

A $k$-partition is a partition in $k$ segments. A predictive
$k$-partition is a $k$-partition in which a model is associated with
each segment, and neighbouring segments have different models. The set
of the predictive $k$-partitions of $\cS$ is denoted $\bP_k$. From
here on, all partitions will be predictive partitions.

Let us call the set of models $D$; for all $d\in D$ we denote by
$\pi_d(i)=\tP(\cs_i|\cS_{i-1},d)$ the probability of the $i$-th letter
given the model $d$ and the previous $i-1$ letters of the sequence.
The likelihood of a segment $\sigma \subset \cS$ given a model $d \in
D$ is the product of the likelihoods of its letters $\tP(\sigma|
d)=\prod_{\cs_i \in \sigma} \pi_d(i)$. For $p$ in $\bP_k$, the
likelihood of $\cS$ given $p$, $\tP(\cS|p,D)$, is the product of the
likelihoods of the predictive segments of $\cS$ defined by the
partition. We have defined a distribution of the likelihoods over
$\bP_k$, $(\tP(\cS|p,D))_{p\in \bP_k}$, and we are looking for the
expectation of this distribution $\tP(\cS|\bP_k,D)=\sum_{p\in \bP_k}
\tP(\cS|\bP_k,D).\tP(p|\bP_k)$.

We denote $m_k(i)$ the expectation of the likelihoods of $\cS_i$ under
the set of the $k$-partitions of $\cS_i$, and $m_k^d(i)$ is the
expectation of the likelihoods of $\cS_i$ under the set of the
$k$-partitions of $\cS_i$ whose model of the last segment is $d$.
These values can be computed with a dynamic programming algorithm (the
demonstration of which is appended):

\hbox{\hspace*{-2em} \vbox{\begin{eqnarray*}  
\forall i \geqslant 0, m_1^d(i)&=&\tP(\cS_i|d) \\
\forall k \geqslant 1, \forall i < k-1, m_k(i) &=& 0\\
\forall k \geqslant 1, \forall i \geqslant k-1, m_k(i) &=& \frac 1{\#D}
\sum_{d\in D} m_k^d(i) \\
\forall k\geqslant 2, \forall i\geqslant k-1, m_k^d(i) &=& \pi_d(i).
\bigg(\frac{i-k+1}{i}.m_k^d(i-1)  \\
&&+ \frac{k-1}{i.(\#D-1)} \left(\#D .m_{k-1}(i-1) -
  m_{k-1}^d(i-1) 
\right) \bigg) 
\end{eqnarray*}
}}

As $\tP(\cS_i|d)$ is the likelihood of a segment given a specific model,
it is computable. We can see that when $i=k-1$, the first term inside
the brackets equals 0, which means that $m_k^d(i)$ can be recursively
computed.

For each $k$, $\tP(\cS|\bP_k,D)=m_k(l-1)$ is the mean likelihood of
$\cS$ under the set of the $k$-predictive partitions.

\medskip

When, in the previous formula, we change $\pi_d(i)$ by
$\pi^{\alpha}_d(i)$, the expectation of the $\alpha$th power of the
likelihood of $\cS$, $E_{p\in \bP_k}(\tP(\cS|p,D)^{\alpha})$, is
computed, which is the $\alpha$-th moment around $0$ of this
distribution. When $\alpha$ is a natural, it is then easy to compute
the $\alpha$-th moment around the mean, such as the variance.

\medskip

This algorithm has a linear time-complexity with the product of the
number of models and the length of the sequence. Hence these
likelihoods are quite computable, even for very long sequences.

\subsection{Estimating the \emph{a posteriori} probabilities}

Considering the segmentation problem, we are actually interested in
the \emph{a posteriori} probability of the number of segments given
the sequence, say $N$. We hypothesize hereafter that the probability
of this number is equal to $\tP(\bP_N|\cS,D)$, even though this
hypothesis deserves a closer examination. However, it is reasonable to
assume that $\tP(N|\cS,D)$ and $\tP(\bP_N|\cS,D)$ have the same modes,
and that a maximal \emph{a posteriori} estimator of $\tP(\bP_N|\cS,D)$
will be a maximal \emph{a posteriori} estimator of $\tP(N|\cS,D)$.

Owing to the bayesian formula $\tP(\bP_N|\cS,D)\propto
\tP(\cS|\bP_N,D) \tP(\bP_N|D)$, an \textit{a priori} on the
distribution of $\tP(\bP_N|D)$ has to be set. If this \textit{a
priori} is uniform with $k$, the \textit{a posteriori} probability is
directly proportional to the likelihood computed in the previous
section: $\tP(\bP_{N=k}|\cS,D)\propto \tP(\cS|\bP_k,D)$.

Another \textit{a priori} is analogous to the HMM modelling: we
consider that the segment length follows a geometrical distribution
with a given mean, say $\lambda$. Then \emph{a priori} $N-1$ follows a
binomial distribution of parameter $\frac \lambda l$, and if we define
a random variable $X \leadsto \text{Bin}(l,\frac \lambda l)$,
$\tP(\bP_{N=k}|\cS,D) \propto \tP(\cS|\bP_k,D).\tP(X=k-1)$.

A more experimental approach is to consider that $\tP(\bP_N|\cS,D)$
follows a given law with some parameters, and to simulate sequences
generated by $k$-partitions to fit at best these parameters,
considering an optimization criterion. An obvious criterion is to
minimize the mean square error of the maximum \emph{a posteriori}
estimation of the numbers of segments.

\subsection{Implementation\label{sec:implementation}}

This algorithm has been implemented in C++, and is freely available
via python modules in Sarment~\citep{Gue05} at the URL:\\
\verb|http://pbil.univ-lyon1.fr/software/sarment/|

The examples of the next section are described in the tutorial at the
same location.

\newpage
\section{Maximum a posteriori estimation\label{sec:testing-applications}}

\subsection{The \textit{a priori} distribution\label{sec:pred-numb-segm}}

To build a good \textit{a posteriori} estimator, we still need to look
for a relevant \textit{a priori} probability on the $\bP_k$. To test
this, I have generated random sequences made up of an alphabet of two
letters (\texttt A and \texttt B), from several Markov models and
random $k$-partitions, for several values of $k$. We denote
$\text{Bern}(\alpha)$ the model where the emission probability of an
\texttt A is $\alpha$ (and that of a \texttt B is $1-\alpha$). The
positions of the limits of the segments were uniformly generated, so
that each segment was at least 50 positions long, and the models were
uniformly assigned to each segment so that no two neighbouring
segments shared the same model. For each $k$, 100 random
$k$-partitions and sequences 10,000 letters in length have thus been
generated. To understand how the algorithm performs on more or less
strongly segmented sequences, the next examples present sequences
generated from models $\text{Bern}(0. 3)$ and $\text{Bern}(0.
7)$, and sequences generated from more similar models
$\text{Bern}(0. 4)$ and $\text{Bern}(0. 6)$. The same models
have been used to compute the likelihoods.

\begin{figure}[p]
  \includegraphics[height=5cm,width=\textwidth]{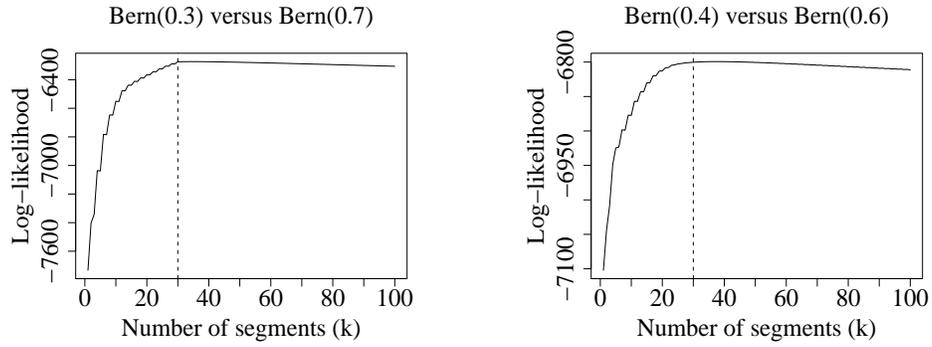}

  \caption{Log-likelihood of two random sequences generated by 30
    segments from two models. The dashed vertical line represents
    $30$ segments.}
  \label{fig:prob_nb_cl}
\end{figure}

\marginpar{Fig.~\ref{fig:prob_nb_cl}}

First, I searched for the number of segments $N$ for which the
sequence has the highest likelihood. It is equivalent to the uniform
\textit{a priori} distribution.

The examples of log-likelihoods in Fig.~\ref{fig:prob_nb_cl} show a
typical behaviour: the neighbourhood of the maximum likelihood can be
reached very quickly, and there are several numbers of segments with a
likelihood ``near'' this maximum. If in the left example, the maximum
is reached on the exact number of segments, this maximum is reached
for a higher number in the right example.

\begin{figure}[p]
  \includegraphics[height=6cm,width=\textwidth]{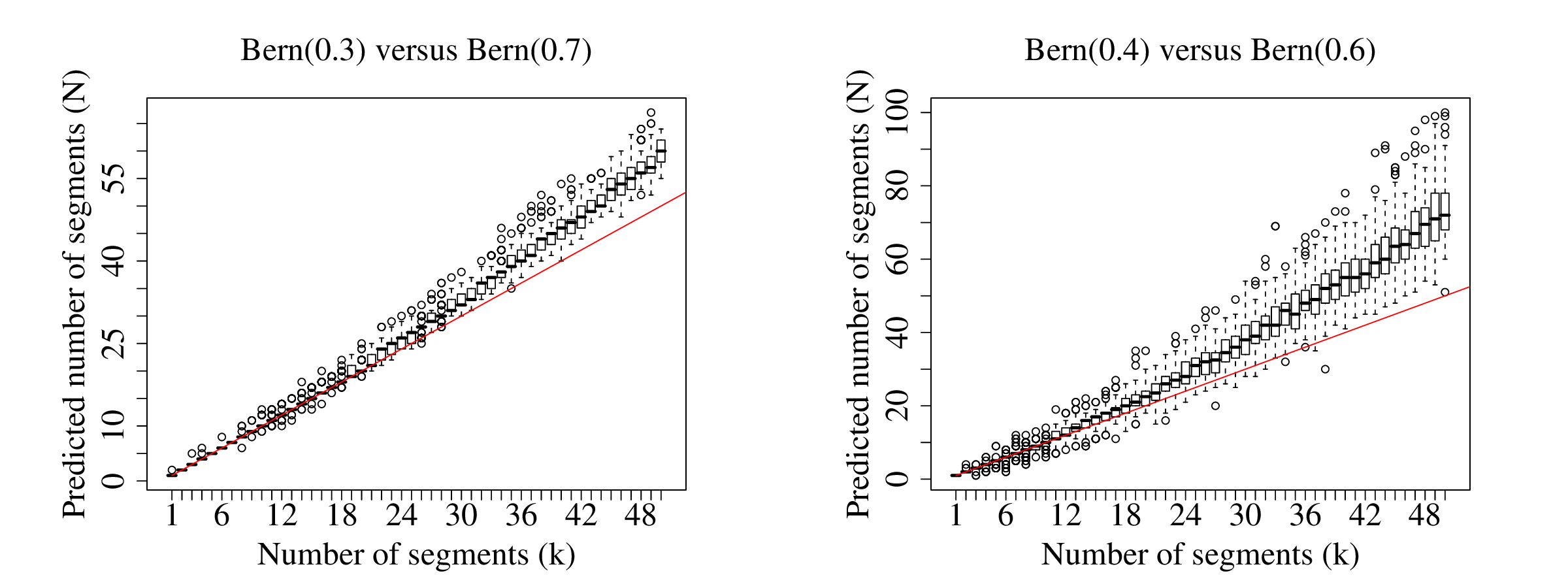}
  \caption{Boxplots of the numbers of segments $N$ reaching the
    maximum likelihood of the sequence, for a simulated number of
    segments $k$ between 1 and 50.
The oblique line represents the right number of segments ($N=k$).
  }
  \label{fig:nb_cl}
\end{figure}

\marginpar{Fig.~\ref{fig:nb_cl}}

Actually, overall, the predicted numbers of segments are in accordance
with the simulated numbers (Fig.~\ref{fig:nb_cl}). However, as the
segments become more difficult to discriminate (when the average size
of the simulated segments decreases or when the models generating the
segments are more similar), the predicted number tends to
over-estimate. This means that the number of segments with the highest
likelihood is not in fact the one most relevant for this prediction,
and another \textit{a priori} than the uniform distribution should be
chosen.


The \textit{a priori} can be based on the length of the segments, as
it is done in HMM modelling. Since in the simulations the
inter-segments positions of the random partitions were uniformly taken
along the sequence, the lengths of the simulated segments followed a
geometric distribution, which should favour the analysis through HMM.

\begin{figure}[p]
  \includegraphics[height=12cm,width=\textwidth]{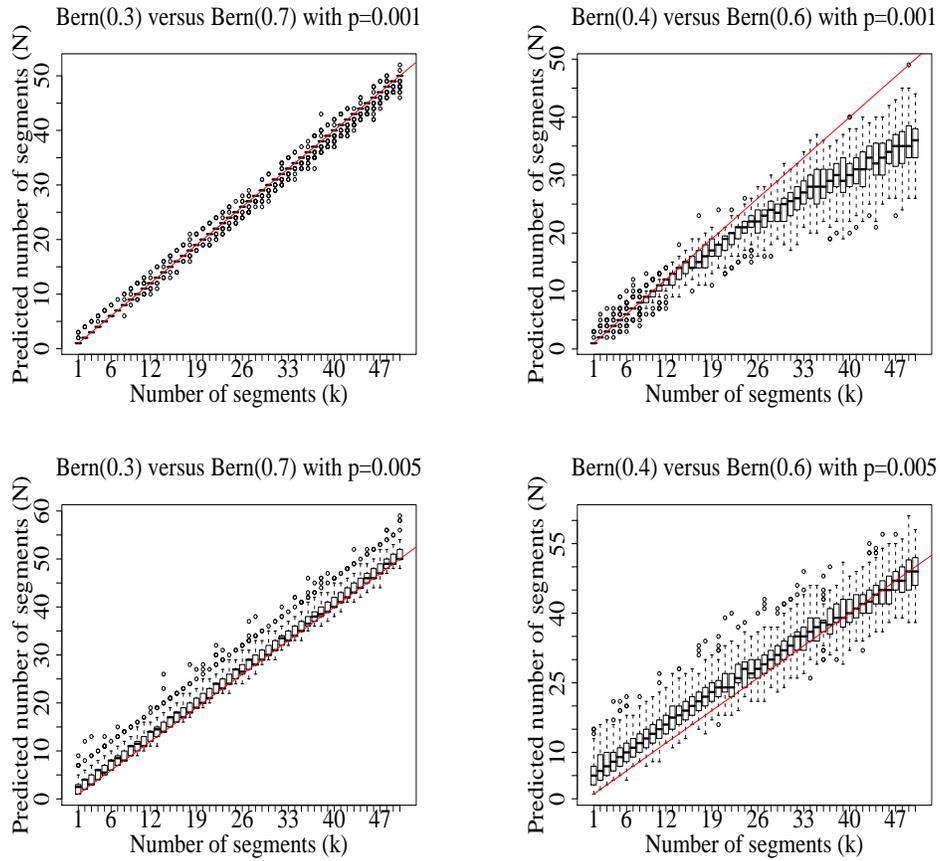}
 \caption{Boxplots of the numbers of segments $N$ predicted by the
    forward-backward algorithm, for a simulated number of segments $k$
    between 1 and 50. The oblique line represents the right number of
    segments ($N=k$). }
  \label{fig:hmm_nb_cl}
\end{figure}

I have studied these sequences with the likelihood algorithm and with
an HMM. The HMM used had the exact Markov models and an additional
parameter $p$ on the probability transitions between the states, so
that the average length of the segments is $1/p$. To get the resulting
partition, I have applied the forward-backward algorithm on the
sequences and successive positions were clustered in a segment when
their most likely state was identical. Since the sequences were 10,000
letters long, the number of predicted segments minus one follows the
binomial law $\text{Bin}(9999,p)$. I used $p=0. 001$ (10 segments)
and $p=0. 005$ (50 segments), and again models $\text{Bern}(0.
3)$ versus $\text{Bern}(0. 7)$ and $\text{Bern}(0. 4)$ versus
$\text{Bern}(0. 6)$ (Fig.~\ref{fig:hmm_nb_cl}).

\marginpar{Fig.~\ref{fig:hmm_nb_cl}}

Figure~\ref{fig:hmm_nb_cl} shows that when the models are distant
($\text{Bern}(0. 3)$ versus $\text{Bern}(0. 7)$), the
forward-backward algorithm performs rather well. However, with
$p=0. 005$ the number of segments is more over-estimated than with
$p=0. 001$, since it tends to increase the number of segments.
When the models are less different, as with $\text{Bern}(0. 4)$
versus $\text{Bern}(0. 6)$, the influence of $p$ becomes critical.
In this example, $p=0. 001$ under-estimates the number of segments
when the real number is over $10$, since this parameter means that
\textit{a priori} on average the sequence has $10$ segments. With
$p=0. 005$ the predictions over-estimate slightly for small
numbers of segments, and they tend to under-estimate as the real
number increases.

We can see that when this estimator is biased, the bias depends on the
value of the inter-state probability and on the real number of
segments in the sequence, which is not known beforehand.

\begin{figure}[p]
  \includegraphics[height=12cm,width=\textwidth]{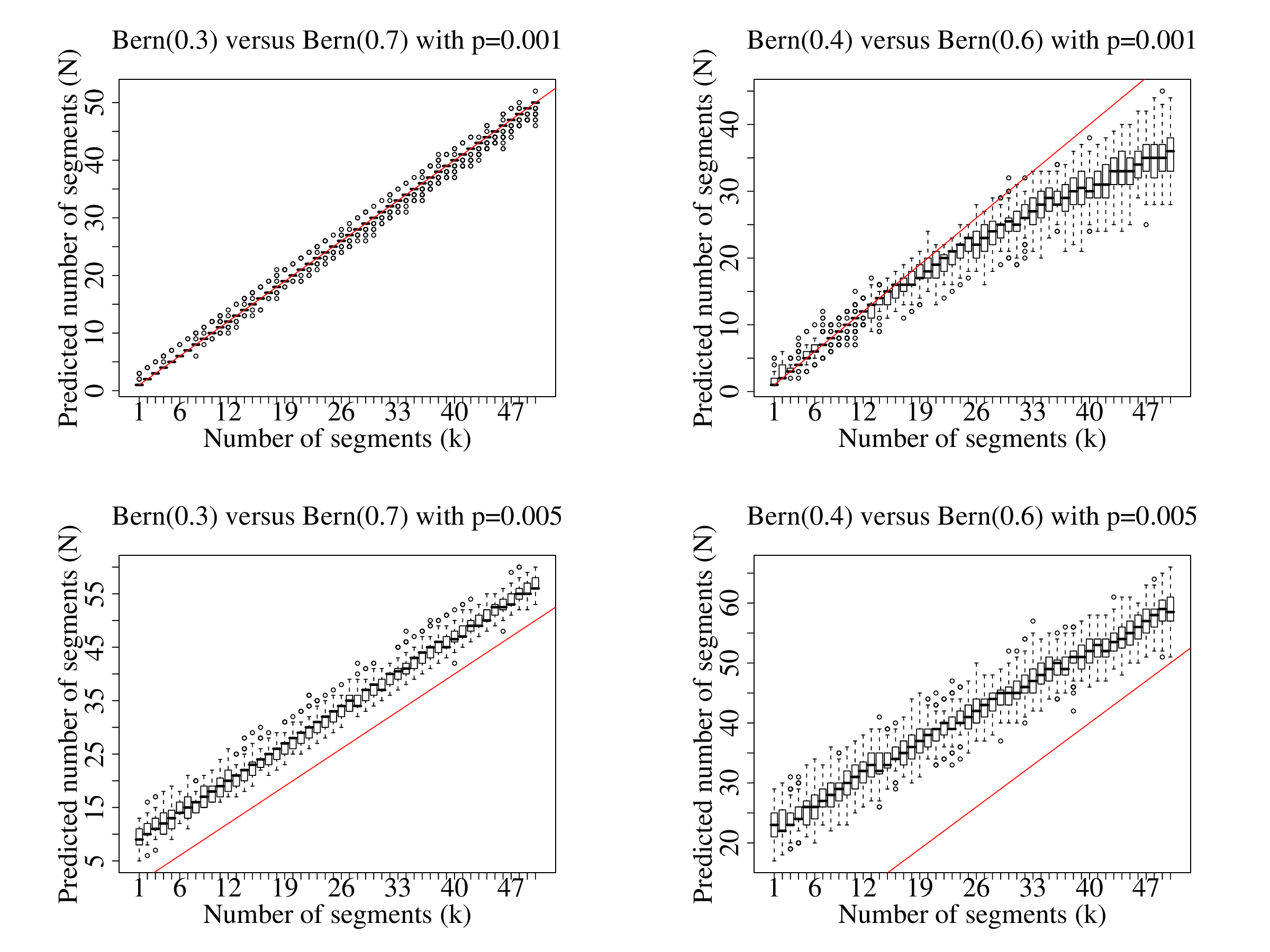}
 \caption{Boxplots of the numbers of segments $N$ with the maximum
    \textit{a posteriori} probability, with a binomial \textit{a
    priori} distribution on $\tP(\bP_N|D)$, for a simulated number of
    segments $k$ between 1 and 50. The oblique line represents the
    right number of segments ($N=k$).}
  \label{fig:bin_nb_cl}
\end{figure}

\marginpar{Fig.~\ref{fig:bin_nb_cl}}

Nonetheless, to study the effect of this \textit{a priori} on the
maximum \textit{a posteriori} estimator, I have set the same binomial
\textit{a priori} distribution on $\tP(\bP_N|D)$, with $p=0. 001$
and $p=0. 005$. We can see in Fig.~\ref{fig:bin_nb_cl} the same
behaviour as with the HMM modelling, but with a much more important
over-estimation of the number of segments when $p=0. 005$. It
means that the tendency of this \textit{a priori} to ``drag'' the
maximum \textit{a posteriori} towards 50 segments is here more
influential. When the real number of segments is near 50, the
over-estimation is lower than in Fig.~\ref{fig:nb_cl}, for the same
reason. Then, even though it corresponds to the modelling of HMM, a
binomial \textit{a priori} is not relevant for maximum \textit{a
posteriori} estimation of the number of segments.

\label{sec:exper-text-priori}

An experimental way to set up an \emph{a priori} distribution is to
define it through a set of parameters, that will be optimized by
simulations. The optimization function is the minimization of the mean
square error between the maximum \emph{a posteriori} estimation and
the real numbers $k$ of segments, summed for all $k$ from 1 to 50.

\begin{figure}[p]
  \includegraphics[height=6cm,width=\textwidth]{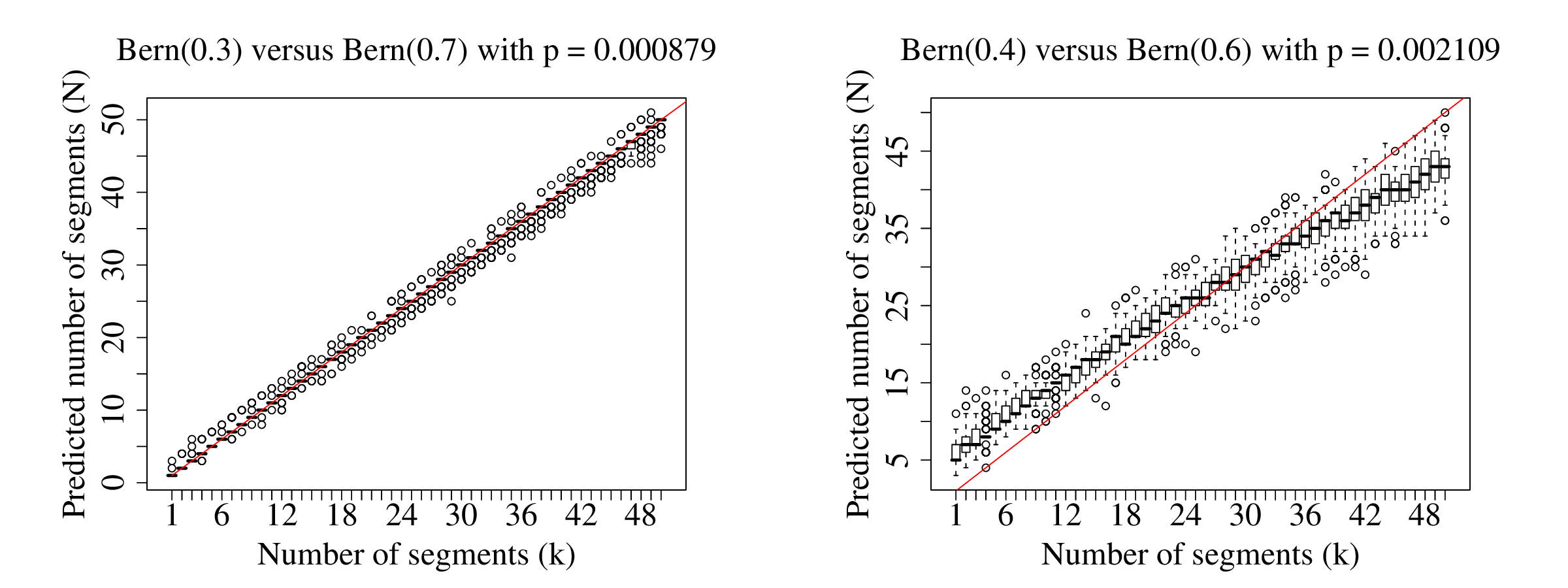}
 \caption{Boxplots of the numbers of segments $N$ with the maximum
    \textit{a posteriori} probability, with a binomial \textit{a
    priori} distribution on $\tP(\bP_N|D)$ and an optimized parameter
    $p$, for a simulated number of segments $k$ between 1 and 50. The
    oblique line represents the right number of segments ($N=k$). }
  \label{fig:optim_bin_nb_cl} 
\end{figure}

A first way would be to optimize the parameter of the binomial
\textit{a priori} distribution. Indeed, the poor efficiency of these
examples could be due to a bad parameter value. In these simulations,
the optimal value $p$ is $0. 00098$ (resp. $0. 0021$) for the
models $\text{Bern}(0. 3)$ versus $\text{Bern}(0. 7)$ (resp.
$\text{Bern}(0. 4)$ versus $\text{Bern}(0. 6)$). The first
optimization is quite efficient (Fig.~\ref{fig:optim_bin_nb_cl}) but
when the segments are less different there is an over-estimation of
the number of segments for small $k$, and an under-estimation for
large $k$, as in the previous section. The correct estimations are
around 25 segments, a balance between over-estimating and
under-estimating all the $k$ between 1 and 50. Then even with an
optimization process, a binomial \textit{a priori} does not give an
efficient \textit{a posteriori} estimator.

\marginpar{Fig.~\ref{fig:optim_bin_nb_cl}}

\begin{figure}[p]
  \includegraphics[height=6cm,width=\textwidth]{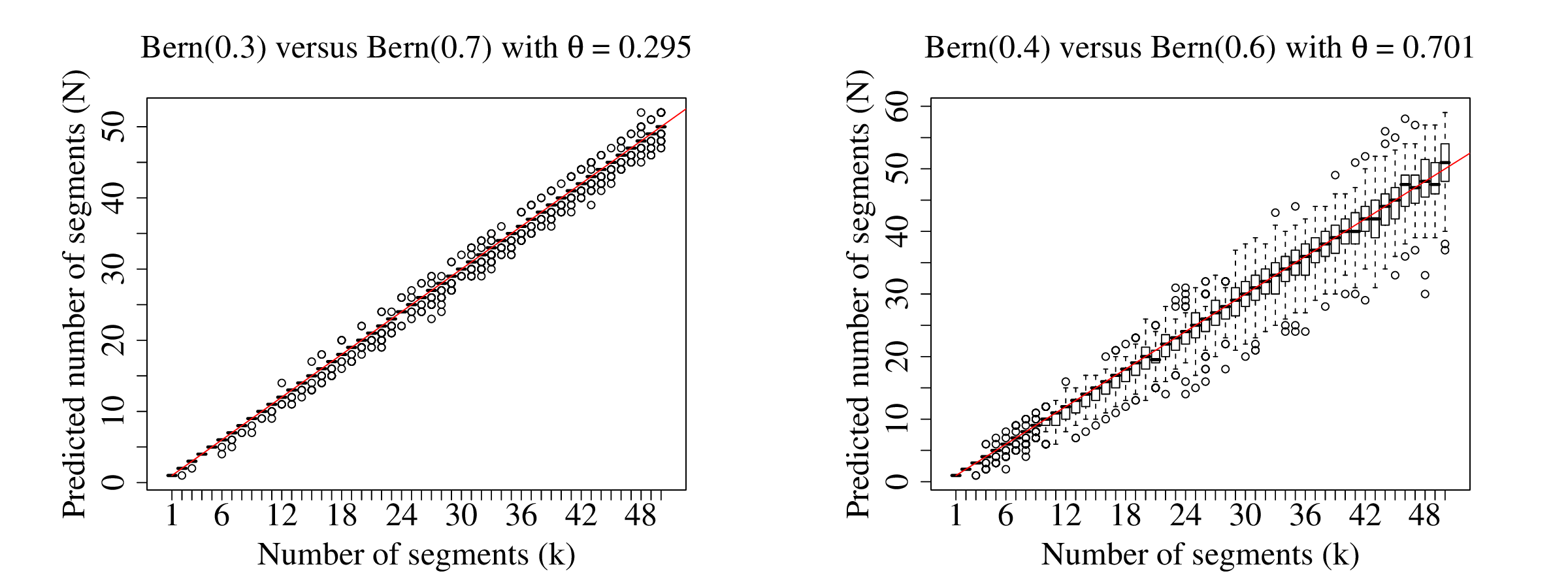}
 \caption{Boxplots of the numbers of segments $N$ with the maximum
    \textit{a posteriori} probability, using an \textit{a priori}
    distribution $\cG(\theta)$ with an optimized $\theta$, for a
    simulated number of segments $k$ between 1 and 50. The oblique
    line represents the right number of segments ($N=k$). }
  \label{fig:optim_lin_nb_cl}
\end{figure}

I tried the same kind of optimization with a geometric \textit{a
priori} distribution $\cG(\theta)$: $\tP(\bP_{N=k}|D,\cS) \propto
\tP(\cS|\bP_k,D). \theta^k$. I have performed twice the same round of
sequence simulations as before, one set for the optimization of the
parameter, and one set to test it on the obtained estimator. On these
examples, when the models are distant enough, as in
$\text{Bern}(0. 3)$ versus $\text{Bern}(0. 7)$, the estimator
is quite accurate, and it is unbiaised, even with $\text{Bern}(0.
4)$ versus $\text{Bern}(0. 6)$ models
(Fig.~\ref{fig:optim_lin_nb_cl}).

\marginpar{Fig.~\ref{fig:optim_lin_nb_cl}}
This example shows that this approach can give good results, even
though it is up to now only experimental. A theoretical study may be
useful to set up an even more efficient \textit{a priori}, and to
prevent the cost of simulations as well as the numerical optimization
process. We can expect this distribution to depend on the set of
models and on the length of the sequence, and it would be quite
interesting to study it thoroughly.

\subsection{CpG islands}

In vertebrate genomes, CpG dinucleotides are mostly methylated and
this methylation entails an hypermutability of these nucleotides, from
CpG to TpG or CpA. A usual measure of this feature is to compute the
ratio of the observed CpG dinucleotides over the expected number when
the nucleotides are independent:
$$\textrm{CpGo/e}=\frac{\textrm{frequency of CpG }}{\textrm{frequency
of C } \times \textrm{frequency of G}}$$

\begin{figure}[p]
  \includegraphics[height=15cm,width=\textwidth]{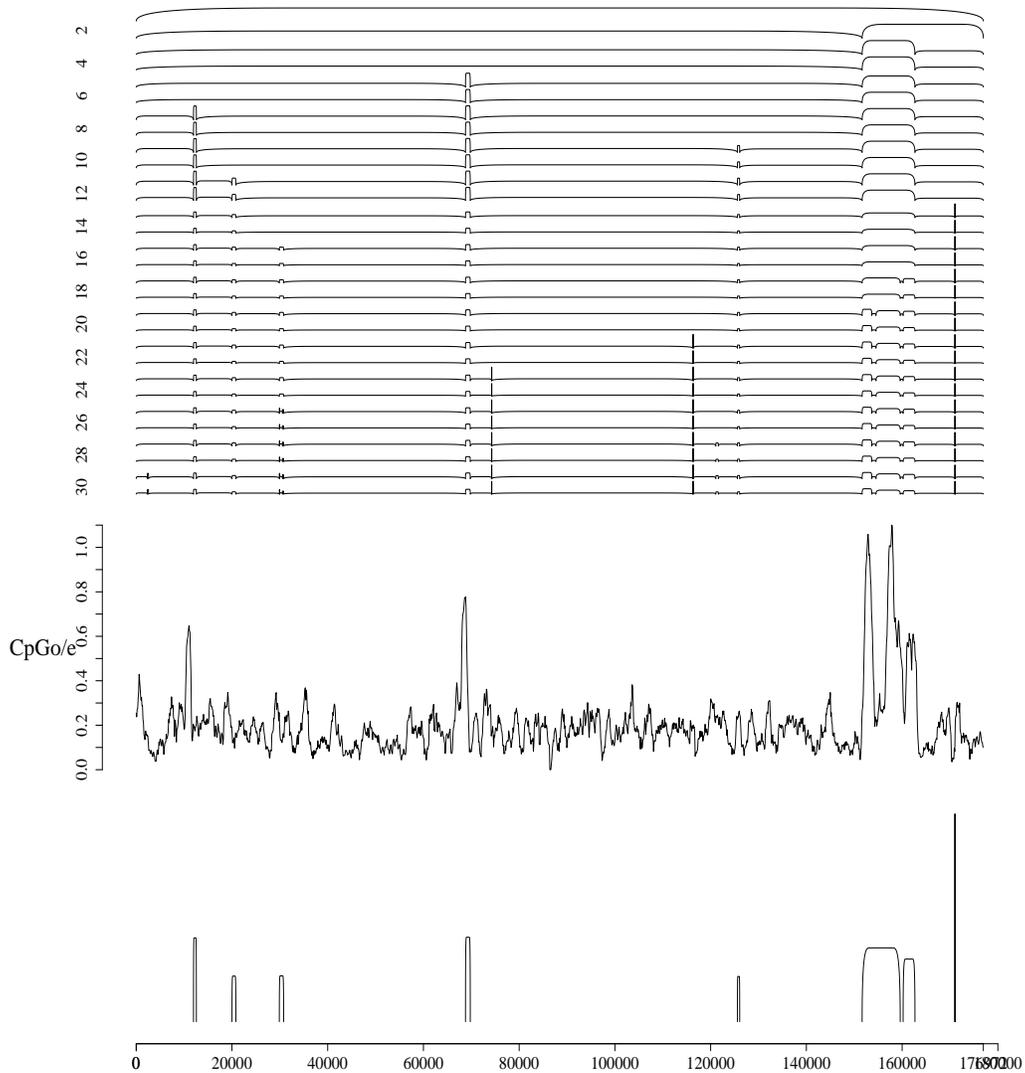}
  \caption{Analysis of the CpG islands from a Mouse genomic sequence.
  The sequence is shown on the x-axis. Top: Partitioning up to 30
  segments. A row of arcs labelled by a number $k$ represents the best
  $k$-partition (only even numbers are shown, for clarity). Each arc
  represents a segment. On each row, the relative height of an arc
  corresponds to the ratio CpGo/e on the segment. Middle: CpGo/e in
  1,000 bases sliding windows. Bottom: Predicted CpG islands of the best
  17-partition of the sequence.}
\label{fig:mus2}
\end{figure}

In some stretches of DNA, known as CpG islands, the CpG dinucleotides
are hypomethylated. These islands are often associated with promoter
regions~\citep{PDM01}. They show a higher CpGo/e than surrounding
sequences, at least $0. 6$. Moreover, a CpG island is expected to
be at least 300 bases long. I wanted to segment a sequence of the
mouse genome to reveal the occurences of CpG islands. The CpGo/e ratio
on this sequence is shown in 1,000 bases sliding windows
(Fig.~\ref{fig:mus2} middle).

\marginpar{Fig.~\ref{fig:mus2}}

\begin{figure}[p]
  \includegraphics[height=6cm,width=\textwidth]{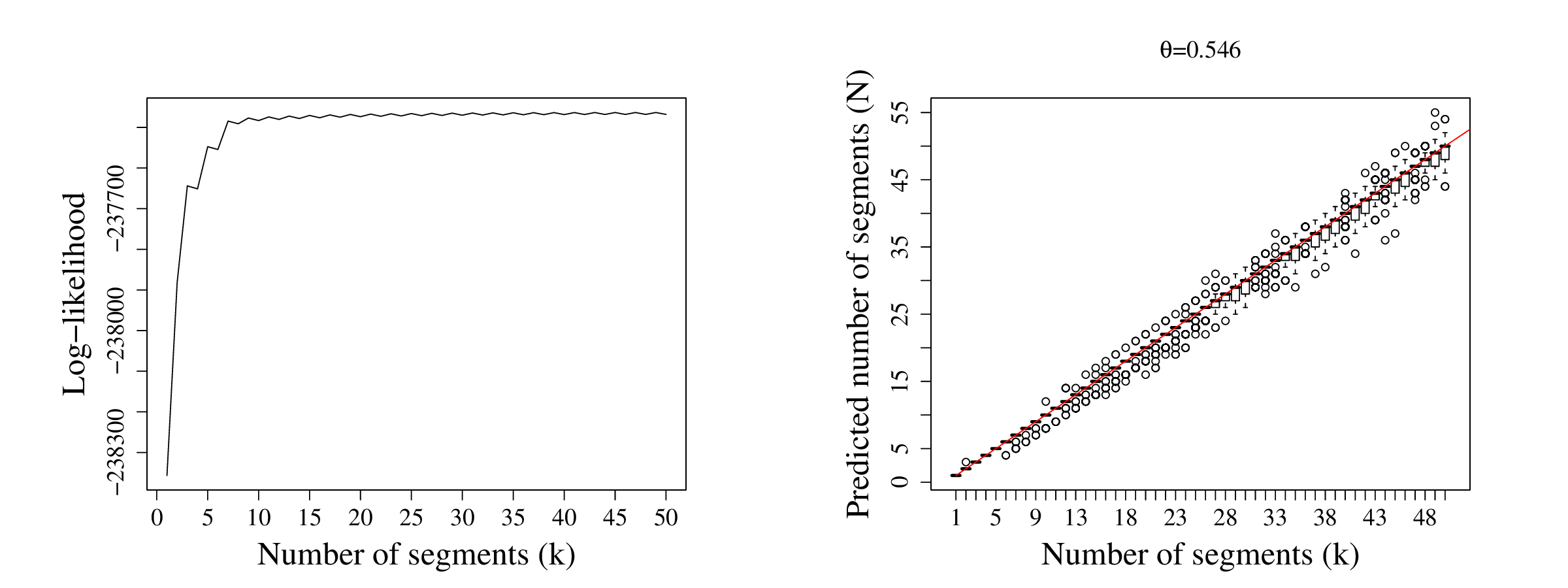}
 \caption{Study of a mouse sequence under the set of the
    $k$-partitions, given the CpG island vs non-Cpg island models.
    Left: Log-likelihood of the sequence, for $k$ numbers of segments,
    with $k$ between 1 and 50. Right: Boxplots of the numbers of
    segments $N$ with the maximum \textit{a posteriori} probability,
    with a geometric \textit{a priori} distribution
    $\cG(0.546)$, for a simulated number of
    segments $k$ between 1 and 50. The simulated sequences were the
    same length than the studied one (176973), and the segments were
    at least 300 long. The oblique line represents the right number of
    segments ($N=k$).}
  \label{fig:mus2_nbe_cl}
\end{figure}

\marginpar{Fig.~\ref{fig:mus2_nbe_cl}}


As described by~\cite{DEKM98}, I defined two first-order Markov
models, built by maximum likelihood on known data: the first is
trained on CpG islands, and the other on segments that are between the
CpG islands. I used those models to compute the segmentation
likelihood on a sequence of the mouse genome
(Fig.~\ref{fig:mus2_nbe_cl}), for up to 50 segments.

I looked for the maximum \textit{a posteriori} estimator of the number
of segments, with a geometric \textit{a priori} distribution, and I
simulated random sequences through the same process as described in
section~\ref{sec:implementation}. The optimization of the maximum
\textit{a posteriori} estimator gives
$\theta=0.546$, and the result of this
optimization is shown in Fig.~\ref{fig:mus2_nbe_cl}. We can see that
this estimator is still unbiased until 50 segments, and quite precise.

With this \textit{a priori}, the maximum \textit{a posteriori}
estimator on the mouse sequence gives 17 segments, and CpG-islands
predicted in the most likely $17$-partition are shown in
Fig.~\ref{fig:mus2} bottom.

\newpage

\section{Discussion}

In this article, I propose an algorithm to compute the moments of the
likelihood of segmentation of a sequence in a number of segments,
given a set of Markov models. This algorithm has a time-complexity
linear with the length of the sequence and the number of models, and
it can be used on very long sequences.

From this likelihood, it should be possible to compare the numbers of
segments to partition a sequence, either through statistical tests or
through a bayesian approach. In a bayesian approach, the \textit{a
priori} distribution of the numbers of classes must be defined, and I
give some examples where a geometric \textit{a priori} distribution
gives a quite precise maximum \textit{a posteriori} estimator. This
has been only validated with simulations, and a full theorerical study
is yet to be undertaken on the \textit{a priori} distribution.
Moreover, it would be quite interesting to define some statistical
tests to assess the relative significance -- confidence intervals and
p-values -- of the numbers of segments, given the models and the
sequence. The fact that the moments of the distribution of the
likelihood can be computed could be useful for this, as well as for an
improvement of the previous estimator.

This algorithm does not put any constraint on the succession of
models, but works as if the transition graph between the models were a
clique. It is easy to see from the Appendix that it can be adapted to
any kind of transition graph, which means that it may be useful in the
context of HMM analysis, for example to check -- or determine -- the
inter-model probabilities of the models, given a sequence. In this
context, it could also be interesting to use the likelihood to enhance
the efficiency of methods related to HMM modelling, for example for
post-analysis of forward-backward algorithm. As in HMM modelling, one
aim would be to compute the probability that a position is predicted
by a model, given a set of models, and possibly given a number of
segments. If the MPP algorithm is equivalent to the Viterbi algorithm
for HMM, computing this probability would be the equivalent of the
forward-backward algorithm in this context.

Even if model inference is out of the topic of this article, it is a
very important feature in sequence analysis, and it will be
interesting to use the likelihood for this. In~\cite{Pol07}, there is
an example of inference of Markov models from a sequence, out of the
context of HMM, but it is practically limited with the numbers of
segments in the sequence and, since it uses the maximum likelihood, an
additional penalization criterion (AIC or BIC) is necessary to handle
this number. It should be possible to use the calculation of the
average likelihood to get rid of these problems. Another inference
process is the maximization, among a set of models, of the average
likelihood. Moreover, it would be relevant to use the bayesian
approach to estimate and simulate \textit{a posteriori} probabilities
for the parameters of the models, given the sequence.

Finally, as I said in the introduction, to my knowledge multi-level
segmentations of sequences are not used for sequence analysis,
although its relevance. An important barrier to this is the lack of
evaluation criteria for these levels. Computing the likelihood for the
successive numbers of segments may then be a quite useful tool to
develop this kind of methodology. It would bring out a much richer
modelling of the sequence.

\section*{Acknowledgment}

The numerous simulations have been made at the PRABI and at the IN2P3
computer center. I thank Meg Woolfit and Daniel Kahn for their useful
corrections.

\section*{Disclosure Statement}

No competing financial interests exist.

\appendix
\section*{Appendix}

Here is a demonstration of the formula described in
section~\ref{sec:method}, keeping the same notations:

We define
\begin{itemize}
\item[] $\bP_k(i)$ the set of the $k$-partitions of $\cS_i$
\item[] $\bP_k^d(i)$ the set of the $k$-partitions of $\cS_i$ whose
  model of the last segment is $d$.
\end{itemize}

$m_k(i)$ is the likelihood of $\cS_i$ under $\bP_k(i)$, $\forall k \geqslant 0, \forall i \geqslant k-1, m_k(i)=\tP(\cS_i
|\bP_k(i))$, and $m^d_k(i)$ is the likelihood of $\cS(i)$ under
$\bP_k^d(i)$,  $\forall k \geqslant 0, \forall i \geqslant k-1,
m^d_k(i)=\tP(\cS_i |\bP^d_k(i))$

If the \textit{a priori} on the last model $d$ is uniform:
\begin{eqnarray}
m_k(i) &=&\tP(\cS_i|\bP_k(i)) = \sum_{d\in D}
    m_k^d(i).\tP(\bP_k^d(i)|\bP_k(i)) \notag \\ 
&=&\frac 1{\#D} \sum_{d\in D} m_k^d(i)\label{eq:3}
\end{eqnarray}

We follow a bayesian approach, in which, for each $k$, all the
$k$-partitions are equiprobable in $\bP_k$.

If we note $d_p(i)$ the model used in partition $p$ at position $i$,
we have for all $k \geqslant 2$ and $ i \geqslant k-1,$
\begin{eqnarray}  
m_k^d(i) &=& \tP(\cS_i|\bP_k^d(i)) \notag \\
&=& \sum_{p \in P_k^d(i)} \tP(\cS_i|p).\tP(p|\bP_k^d(i))\notag \\
&=& \sum_{p \in P_k^d(i)} \tP(\cS_i|p).\#\bP_k^d(i)^{-1}\notag \\
&=& \sum_{\substack{p \in \bP_k^d(i) \\ d_p(i-1)=d}}
\tP(\cS_i|p).\#\bP_k^d(i)^{-1}
+ \sum_{\substack{p \in \bP_k^d(i) \\  d_p(i-1) \neq d}} 
\tP(\cS_i|p).\#\bP_k^d(i)^{-1} \notag 
\end{eqnarray}

If $p\in \bP_k^d(i)$ and $d_p(i-1) = d$, $p$ is like a
$k$-partition $p'$ of $\cS_{i-1}$ whose last model, $d$, is used to
emit $s_i$. So $\tP(\cS_i|p)=\pi_d(i).\tP(\cS_{i-1}|p')$ with $p'\in
\bP_k^d(i-1)$.

If $p\in \bP_k^d(i)$ and $d_p(i-1) \neq d$, $p$ is like a
$k-1$-partition $p'$ of $\cS_{i-1}$ whose last model, $d'$, is
different from $d$. So $\tP(\cS_i|p)=\pi_d(i).\tP(\cS_{i-1}|p')$ with
$p'\in \bP_{k-1}^{d'}(i-1)$.

Hence
\begin{eqnarray}  
m_k^d(i) &=& \pi_d(i).\left(\sum_{p \in \bP_k^d(i-1)}
\tP(\cS_{i-1}|p).\#\bP_k^d(i)^{-1} \right. \notag \\
\lefteqn{\left. \qquad  \qquad + \sum_{d' \neq d} \sum_{p \in \bP_{k-1}^{d'}(i-1)}
\tP(\cS_{i-1}|p).\#\bP_k^d(i)^{-1} \right)}\label{eq:1} \
\end{eqnarray}
In a partition of $\bP_k^d(i)$, the last model is $d$, the one before
any of the $\#D-1$ other ones, and so on for the $k-2$ remaining
models. So there are $(\#D-1)^{k-1}$ possible sets of models
for this partition. Moreover, the limits of the segments are defined
by $k-1$ positions in the $i$ possible, so there are $\cC_i^{k-1}$
possible sets of positions. So
\begin{eqnarray}
  \#\bP_k^d(i)&=&\cC_i^{k-1} (\#D-1)^{k-1} \notag \\
&=&\frac{i!}{(k-1)!(i-k+1)!} (\#D-1)^{k-1} \notag \\
&=&\frac{i}{i-k+1} \#\bP_k^d(i-1) \notag
\end{eqnarray}
and
\begin{eqnarray}
  \#\bP_k^d(i)&=&\frac{i!}{(k-1)!(i-k+1)!} (\#D-1)^{k-1} \notag \\
&=&\frac{i}{k-1} (\#D-1) \#\bP_{k-1}^d(i-1) \notag
\end{eqnarray}

If we replace $\#\bP_k^d(i)$ in~(\ref{eq:1}):
\begin{eqnarray}  
m_k^d(i) &=& \pi_d(i). \left(\sum_{p \in \bP_k^d(i-1)}
\tP(\cS_{i-1}|p).\frac{i-k+1}{i}\#\bP_k^d(i-1)^{-1} \right. \notag \\
\lefteqn{ \left. \qquad + \sum_{d' \neq d} \sum_{p \in \bP_{k-1}^{d'}(i-1)} 
\tP(\cS_{i-1}|p).\frac{k-1}{i(\#D-1)}\#\bP_{k-1}^d(i-1)^{-1} \right) \notag }\\
&=& \pi_d(i).\left( \frac{i-k+1}{i}\sum_{p \in \bP_k^d(i-1)}
\tP(\cS_{i-1}|p).\tP(p|p\in \bP_k^d(i-1))  \right. \notag \\
\lefteqn{\left.  \qquad + \frac{k-1}{i(\#D-1)} \sum_{d' \neq d} \sum_{p \in 
\bP_{k-1}^{d'}(i-1)} 
\tP(\cS_{i-1}|p).\tP(p|p\in \bP_{k-1}^{d'}(i-1)) \right)} \notag \\
&=& \pi_d(i). \left( \frac{i-k+1}{i}.\tP(\cS_{i-1}|\bP_k^d(i-1))
\vphantom{  \frac{k-1}{i(\#D-1)} \sum_{d' \neq d} \tP(cS_{i-1}|\bP_{k-1}^{d'}(i-1))} \right.
\notag
\\
\lefteqn{\left.  \qquad +
 \frac{k-1}{i(\#D-1)} \sum_{d' \neq d} 
\tP(\cS_{i-1}|\bP_{k-1}^{d'}(i-1)) \right)} \notag\\
m_k^d(i) &=& \pi_d(i). \left(\frac{i-k+1}{i}.m_k^d(i-1)+
 \frac{k-1}{i(\#D-1)} \sum_{d' \neq d} m_{k-1}^{d'}(i-1) \right) \notag
\end{eqnarray}

And to make the algorithm faster, from~~(\ref{eq:3}),
$$\sum_{d' \neq d} m_{k-1}^{d'}(i-1)= \#D . m_{k-1}(i-1) -
m_{k-1}^d(i-1)$$
gives the formula.

\bibliographystyle{apalike}
\bibliography{biblio}


\end{document}